\documentclass{emulateapj} 

\usepackage{color,epsfig} 
\definecolor{brown}{rgb}{0.6,0.4,0.2} 
\definecolor{purple}{rgb}{0.5,0,0.5} 

\shorttitle{Water in Molecular Supernova Remnants} 
\shortauthors{}

\newcommand{\kms}{km~s$^{-1}$}

\newcommand{\spitzer}{\textit{Spitzer}} 

\slugcomment{Accepted for the publication in ApJ, August, 2015}
\shorttitle{Extremely Broad Water Detection in the Supernova Remnant G349.7+0.2}

\begin{document}

\title{Detection of Extremely Broad Water Emission from the molecular cloud interacting Supernova Remnant G349.7+0.2{\footnote {thanks{\it Herschel} is an ESA space observatory with science instruments provided by European-led Principal Investigator consortia and with important participation from NASA}}}

\author{ 
J. Rho\altaffilmark{1, 2},  
J. W. Hewitt\altaffilmark{3}, 
A. Boogert\altaffilmark{4},
M. Kaufman\altaffilmark{5} 
and A. Gusdorf\altaffilmark{6}
} 
\altaffiltext{1}{
SETI Institute, 189 N. Bernardo Ave, 
Mountain View, CA 94043; jrho@seti.org}
\altaffiltext{2}{NASA Ames Research Center, MS 211-1, Moffett Field, CA 94043}
\altaffiltext{3}{CRESST/University of Maryland, Baltimore County, Baltimore, MD 21250 and
NASA Goddard Space Flight Center, Greenbelt, MD 20771, USA; john.w.hewitt@nasa.gov }
\altaffiltext{4}{
SOFIA Science Center, NASA Ames Research Center, MS 211-1,
Moffett Field, CA 94035; aboogert@sofia.usra.edu}
\altaffiltext{5}{Department of Physics \& Astronomy, San Jose State University, San Jose, CA 95192-0106; michael.kaufman@sjsu.edu}
\altaffiltext{6}{LERMA, UMR 8112 du CNRS, Observatoire de Paris, \'{E}cole Normale Supérieure, 24 rue Lhomond, 75231 Paris Cedex 05, France; antoine.gusdorf@lra.ens.fr}

\begin{abstract}

We performed {\it Herschel} HIFI, PACS and SPIRE
observations towards the molecular cloud interacting
supernova remnant G349.7+0.2. An extremely broad emission
line was detected at 557 GHz from the ground state
transition 1$_{10}$-1$_{01}$ of ortho-water. This water line
can be separated into three velocity components with widths
of 144, 27 and 4 km~s$^{-1}$. The 144 km~s$^{-1}$ component
is the broadest water line detected to date in the
literature. This extremely broad line width shows importance
of probing shock dynamics. PACS observations revealed 3
additional ortho-water lines, as well as numerous high-J
carbon monoxide (CO) lines. No para-water lines were
detected. 
The extremely broad water line is indicative of a high velocity
shock, which is supported by the observed CO rotational diagram that
was reproduced with a J-shock model with a density of 10$^4$ cm$^{-3}$
and a shock velocity of 80 km s$^{-1}$. Two far-infrared
fine-structure lines, [O~I] at 145 micron and [C~II] line at 157
micron, are also consistent with the high velocity J-shock model. The
extremely broad water line could be simply from short-lived molecules
that have not been destroyed in high velocity J-shocks; however, it
may be from more complicated geometry such as high-velocity water
bullets or a shell expanding in high velocity. 
We estimate the CO and H$_2$O densities, column densities, and temperatures by 
comparison with RADEX and detailed shock models.


\end{abstract} 



\section{Introduction}

Water is the basic building block of life and is often used as a
diagnostic for forms of life on planets
\citep{archer14}. 
Water may be formed by ion-neutral reactions in cold gas, through grain surface reactions 
or neutral-neutral reactions at higher temperatures (e.g. Hollenbach et al. 2009). 
High water abundances are possible behind shocks either due to chemical  reactions or 
due to sputtering of ice mantles.
One of the most important post-shock reactions is $\rm O+H_2 \Rightarrow$
OH + H  followed by $\rm OH + H_2 \Rightarrow H{_2}O + H$, which
rapidly forms H$_2$O in the gas phase. 

Theoretical models predict that in shocks water is a powerful coolant,
second in importance only to H$_2$ \citep{kaufman96,flower10}.
However, recent observations suggest that H$_2$O may be less important
as a coolant and the water abundance may be reduced in shocks
\citep{karska14,santangelo14}. On the other hand, \citet{neufeld14}
find that H$_2$O formation is as efficient as expected, driving all O
into H$_2$O, as soon as shock speeds are high enough  
to convert O-bearing ices to the gas phase. Therefore, the study of water in shocks
is important to resolve this controversy. 

Water is widely detected in
protostellar outflows \citep[e.g.][and references
therein]{vandishoeck11}. 
In contrast, to date, water has
been detected from only three supernova remnants, namely 3C391, IC 443 and W28
\citep{reach98, snell05, neufeld14}. Of these, only IC 443 exhibited broad water
lines, with line widths in excess of ~30 $\rm km\,s^{-1}$ (FWHM). Broad CO lines
($\sim$20-40 km s$^{-1}$) have been detected from a handful  of SNRs including
3C391 \citep{reach98}, IC443 \citep{vandishoeck93}, W28 and W44 \citep{reach05,
gusdorf12, anderl14}. Despite the increasing number of
SNRs that are known to be interacting with molecular clouds, attempts to model
the observed shock diagnostics revealed in these spectra have mainly indicated
the complexity of interstellar shocks \citep[e.g.][H09 hereafter]{hewitt09}.

In this letter, we present Herschel HIFI spectra of ground-state
emission from ortho-H$_2$O; this line emission, which includes
extremely broad velocity components, is compared with additional
H$_2$O lines, CO lines, and fine structure lines of [O~I] and [C~II],
all detected with the Herschel PACS and SPIRE instruments.
We show that the water lines can probe shock dynamics and chemistry, 
and discuss the physical conditions of water and CO emission.


\section{Observations}

We performed {\it Herschel} HIFI and PACS observations towards the
molecular supernova remnant G349.7+0.2 on board of the Herschel Space
Observatory. The spectra have been reduced using recent version of
Herschel Interactive Processing Environment (HIPE) version 12.1. HIFI
observations of H$_2$O were conducted on 2012 September 29
(obsid=1342251666) and 2012 September 22 (obsid= 1342251494) in
pointing mode with a position switch (offset of -3$'$ in RA and 3$'$
in Dec) towards $17^{\rm h} 18^{\rm m} 00.50^{\rm s}$ and Dec.\
$-37^\circ$26$^{\prime} 35^{\prime \prime}$ (J2000). Integration times
were 915 s in Band 1a (557 GHz), and 1936 s in Band 4a (1113 GHz). The
spectral resolutions are 1.10 MHz and 0.25 MHz for HIFI WBS and HRS,
respectively. Figure 1 shows the field of view (FOV) of the PACS and
HIFI instruments. The ground state ortho-water line, 1$_{10}$-1$_{01}$
transition, at 557 GHz was detected, as shown in Figure 2, while no
line was detected at the frequency of the  para-H$_2$O line at 1113
GHz. We examined the off-position spectra at 557 GHz and concluded
that the position is clean and does not contaminate the observed
emission. 
The main beam efficiencies were corrected
based on the release note of new beam efficiency measured on 
Mars{\footnote{http://herschel.esac.esa.int}} which results in
22\% increase in the line brightnesses for 557 GHz line compared to
that using \citet{roelfsema12}. The beam sizes of the 557 and 1113
GHz water line observations are 44$''$ and 19$''$,
respectively.  We have used spectral settings that avoid spurs in
Bands 1a and 4a. 

PACS spectral observations took place on 2012 October 5 for 3,498 s
(obsid = 1342252274), and 
covered from 72.2 to 90.76 $\mu$m and from 112.2 to 181.4
$\mu$m. The pointed observation was made in chopping/nodding Mode
with medium chopping throw. The PACS FOV is overlapped with
the HIFI FOV except for one pixel on the top right where an 
ultra-compact HII region (UC HII) is present (see Fig.
\ref{g349images}a). This pixel corresponds to the brightest pixel on
the PACS continuum map. We extracted PACS spectra at the same region
covered by HIFI which excluded the brightest pixel. 
The PACS observations yielded a 5$\times$5 grid map with
a pixel size of 9.4$''$$\times$9.4$''$. The spectral resolution ranges
from 1000 at 200 $\mu$m to 5000 at 51$\mu$m. SPIRE FTS observations
took place on 2012 September 24 for 3,003 s (obsid = 1342251325). The
beam size of the SPIRE SSW and SLW are 17-21$''$ and 29$''$-42$''$,
respectively.

\section{Broad Water Line with HIFI}

Figure \ref{waterhifispec} shows the broad 557 GHz water line observed
towards the SNR G349.7+0.2. The red branch shows prominent wings with
a smooth Gaussian profile while the blue branch does not and instead
features some dips in the wings which are likely absorption lines in
the line of sight. We have fit the spectrum with Gaussian lines and
Table \ref{tablehifi} summarizes the widths (FWHM) of velocity
($\Delta$V), the velocity at center (V$_o$), the integrated intensity
and the surface brightness (in nWm$^{-2}$sr$^{-1}$ = 10$^{-9}$Wm$^{-2}$sr$^{-1}$). 
The spectrum can be reproduced by three
kinematic components with widths of 144, 27 and 4
km~s$^{-1}$. respectively; we refer to them as extremely broad, broad
and narrow water lines (EBWL, BWL, and NWL).
G349.7+0.2 is the only SNR which reveals this extremely broad (144 \kms)
component among the molecular cloud interacting SNRs we observed
with Herschel.
The LSR velocity of G349.7+0.2 is 16 \kms \citep{dubner04} and the
velocity shifts of the three components are red-shifted at +18, +3,
and +1.4 \kms\ for EBWL, BWL, and NWL, respectively. This water line
exhibits shock dynamics revealed by water.


The X-ray temperature observed for G349.7+0.2 is 0.7-1$\times$10$^7$K
\citep{lazendic05}, which implies a shock velocity of 760 km s$^{-1}$.
When assuming pressure equilibrium between the inter-cloud medium
(ICM) and clouds, and assuming a ICM density (n$_{ICM}$) of $\sim$5-10
cm$^{-3}$ from X-ray gas \citep{lazendic05,chevalier99}, the shock
velocity in the cloud (V$_{sc}$ = $\sqrt {(n_o/n_c)}$$\times$V$_s$) is
approximately 55-76 and 17-24 \kms\ for n$_{c}$ of 10$^{3}$ and and
10$^{4}$ cm$^{-3}$, respectively. 
The observed velocity of 144  \kms\ is much higher that the expected
shock velocity in a cloud when a pressure equilibrium is assumed.

The velocity component of 144 \kms\ is due to strong supernova shocks
and the broadest water line detected to date in the literature. This
may be compared with widths of 81 \kms\ \citep{mottram15}, 50 \kms\
\citep{leurini14}, 30 \kms\ \citep{melnick10} and 20 \kms\
\citep{kama13} from low-mass protostars, high-mass star-forming
regions, and objects towards Orion and OMC-2 FIR 4, respectively. Similar
widths ($\sim$50-60 \kms) are associated with high velocity water
bullets in low-mass protostars \citep{kristensen11}.
The EBW emission could arise from the
high-velocity shell structure seen in the H$_2$ emission with components
covering the velocity range -40 \kms\ to +40 \kms \citep{lazendic10},
while a similar structure could be giving rise to the broad water emission.


The HIFI water spectrum of G349.7+0.2 shows another interesting
feature, different from other SNRs (e.g. S05) or astronomical objects: a narrow
(FWHM=3.7 km/s) emission line (the NWL component). 
Narrow water lines
are seen toward low mass (Kristensen et al. 2012) and high-mass YSOs
(Marseille et al. 2010). They often have P Cygni and reversed P Cygni
profiles, originating from envelope expansion and infall,
respectively. In some cases only narrow emission lines are observed,
in particular for YSOs associated with PDRs (Kristensen et al. 2012),
and this gas may have been released from ices by photodesorption. For
G349.7+0.2, the existence of ices along the line of sight is expected following 
the extinction of $A_v$ = 30 mag, high densities and temperatures
below the sublimation temperature of $\sim$100 K. 
This indicates there is a source of non-thermal desorption, particularly
due to photodesorption \citep{hollenbach09, caselli12a,caselli12b}.
The NWL might trace sublimation or photodesorption
of the ices near the UC HII region, although it is presently unclear if
the UC HII region is associated with the SNR, or if its location is related to the dense cloud
which interacts with the SNR.  
Alternately, a strong cosmic ray flux or a secondary
FUV field generated by the cosmic ray destruction of H$_2$ might
photodesorb H$_2$O, but in this case a strongly enhanced cosmic ray
flux would be needed, considering that the observed NWL is ten times
stronger than that observed in the dense core L1544 (Caselli et
al. 2012). 

More likely, however, shocks may have released H$_2$O into
the gas phase.  The narrow component of CO showed blue- or red-shifted
emission and the central velocity differs at different positions; the
CO gas is suggested to have shocked origin (Dubner et al. 2004). The
fact that the velocity of NWL is consistent with the CO indicates that they are
from the same gas. We suggest that the NWL is likely shocked gas from
denser clumps or reformed molecules behind a dissociative shock (see Section 5
for details). Further observational evidence would be needed to
determine the origin of the NWL. The BWL component with FWHM of 27 \kms\ 
is comparable to those in IC 443 observed by Snell et
al. (2005). The width of the component is consistent with the C-shock
that is responsible for H$_2$ emission (Hewitt et al. 2009).

\section{Molecular and Fine-structure lines with PACS and SPIRE}

\begin{figure}
\plotone{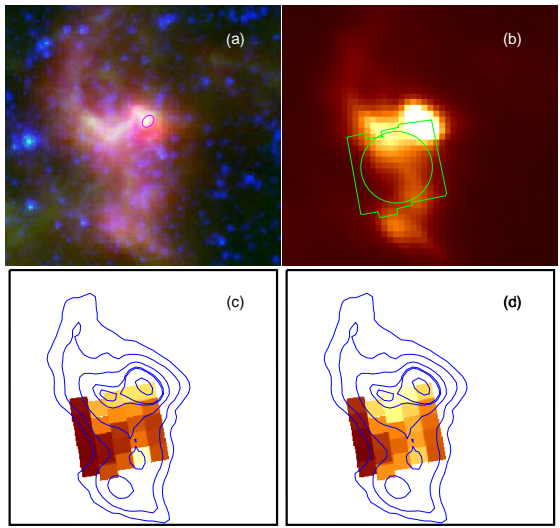}
\caption{{\it Herschel} and {\it Spitzer} images of G347.9+0.2. 
(a) {\it Spitzer} three color images (blue, green and red represents 4.5, 8 and 24
$\mu$m, respectively). MM(1) source which is a possible UC HII region
\citep[see][]{lazendic10} is marked as an ellipse in red. (b) Herschel
PACS broad band at 70 $\mu$m where the brightest peak (right side) is
MM(1) source and the second strongest peak (left side) is center of
the SNR. The FOVs of HIFI (a circle: 557 GHz line) and PACS (a square)
are marked. (c) Water map at 113 $\mu$m (c) and (d) [O~I] map at 145
$\mu$m generated by PACS spectral cube. The contours are from the PACS
70 $\mu$m map. The contour levels correspond to 0.24,0.37, 0.72,1.84, 1.96 and 
5.54 Jy/(3.2$''$x3.2$''$ pixel). 
The maps have 3 arcmin FOV centered on $17^{\rm h}
18^{\rm m} 00^{\rm s}$ and Dec.\ $-37^\circ$26$^{\prime} 20.93^{\prime
\prime}$ (J2000).}
\label{g349images}
\end{figure}

\begin{figure}
\plotone{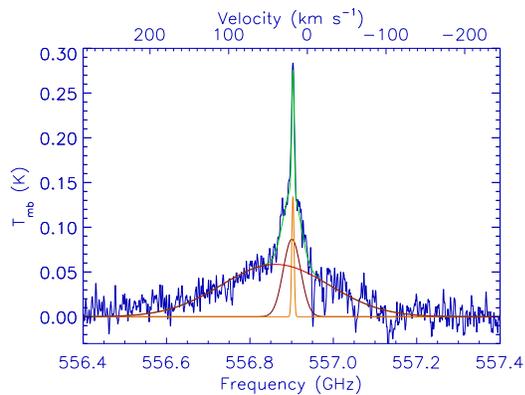}
\caption{{\it Herschel HIFI} spectrum of G349.7+0.2 shows a broad water line at 557
GHz. The gaussian fits (total fit is marked in green) reveal the width
of three kinematic components of 144 (in red), 27 (in brown)  and 4
(in orange) km s$^{-1}$.
}
\label{waterhifispec}
\end{figure}

\begin{figure}
\plotone{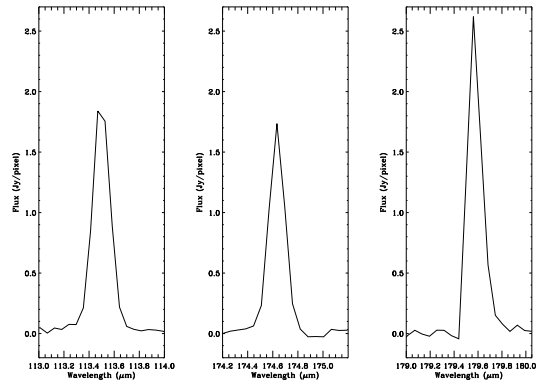}
\caption{Three {\it Herschel PACS} water spectra of G349.7+0.2. The lines are unresolved.
}
\label{waterpacsspec}
\end{figure}

The H$_2$O, CO, [OI] and [CII] lines detected using PACS and SPIRE are 
listed in Table \ref{tablepacs}. The H$_2$O lines detected with PACS
are shown in Figure \ref{waterpacsspec}. (We also detected OH lines; these
will be discussed in a future paper.) 
Using the PACS cube, we produced three line maps of
H$_2$O, CO and [O~I]. 
The spatial distribution of the
H$_2$O and [O~I] lines are shown in Figure \ref{g349images}. H$_2$O emission is
strongest at the southern shell of the SNR, while [O~I] is strongest toward
the SNR central position. CO lines from J$_{upper}$=4 to 13 were detected with SPIRE,
and most CO lines from J$_{upper}$=16 to 36 were detected with PACS. Detected
ortho-H$_2$O lines are seen at 1670 (179.6), 1717 (174.7), 2640 (113.6) GHz
($\mu$m). All ortho-water lines that are within our wavelength range were
detected, except 1097 GHz (273.475 $\mu$m) for which we only note a hint of
emission in the SPIRE data. 
The CO map (not shown) is very similar to the H$_2$O map, and 
the H$_2$O map is globally similar to the [O~I]
map. The H$_2$O peak is between the two bright peaks of the PACS
70 $\mu$m map, while the peak [O I] emission is close to the left bright peak
of the PACS 70 $\mu$m map. None of the para-H$_2$O lines were detected including
the ground state para-H$_2$O 1$_{11}$-O$_{00}$ line at 1113 GHz observed with
HIFI. The upper limit of the 1113 GHz line is 0.06 nWm$^{-2}$ sr$^{-1}$, which is a factor
of 15 fainter than the sensitivity of 557 GHz line.

\section{Discussion}

In order to understand the physical conditions in the shocked gas, we
begin by constructing an excitation diagram for the observed CO lines.
A two-temperature fit yields a low temperature (T$_{\rm low}$) of 170
K with a column density (N$_{\rm low}$) of 1$\times$10$^{16}$
cm$^{-2}$, and a high temperature (T$_{high}$) of 560 K with (N$_{\rm
high}$) of 1$\times$10$^{15}$ cm$^{-2}$. These may be compared with
the two-temperature LTE fit of H$_2$ emission observed with {\it Spitzer} by
\citet{hewitt09} who have found a warm component of $N(H_2)_{warm}$ =
2.8$\times 10^{20}$cm$^{-2}$, and T$_{warm}$ = 467 K and OPR$_{warm}$
= 1, and N($H_2)_{hot}$ = 5.2$\times 10^{18}$cm$^{-2}$, and T$_{hot}$
= 1647 K. Comparing the CO and H$_2$ column densities in the two
components gives CO abundances $x({\rm CO})\sim 3\times
10^{-5}-2\times 10^{-4}$. The CO rotational diagram as a function of
J$_{upper}$ are shown in Figure \ref{coshockmodel}. The surface
brightness of CO peaks at approximately J$_{upper}$ of 15.

\begin{figure}
\plotone{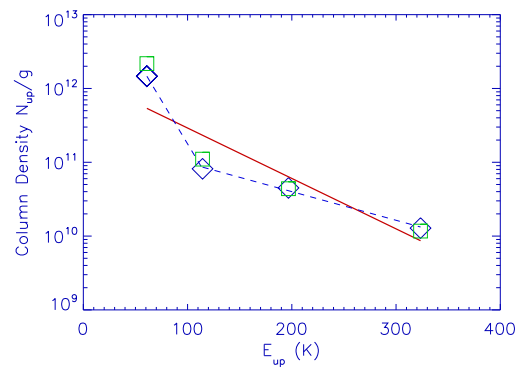}
\caption{Water rotational diagram using the HIFI (E$_{up}$ of 61 K using the intensity
integrated over the whole velocity profile) and PACS data. Only ortho-H$_2$O
lines were detected. The one-temperature fit (in red) indicates N(H$_2$O) =
1.4$\times$10$^{12}$ cm$^{-2}$ and a temperature of 68 K. The two-temperature
LTE fit marked in blue yielded a low temperature (T$_{\rm low}$) of 18 K with a
column density (N$_{\rm low}$) of 4.0$\times$10$^{13}$ cm$^{-2}$, and a high
temperature (T$_{high}$) of 112 K with (N$_{\rm high}$) of 2.3$\times$10$^{11}$
cm$^{-2}$. Non-LTE models (n=10$^5$ cm$^{-3}$ and N$_{H_2O}$ = 3$\times$10$^{13}$ cm$^{-2}$
with a range of kinetic temperature of 100 - 1000) are marked as squares in green.
}
\label{waterrotdiag}
\end{figure}

\begin{figure}[!h]
\plotone{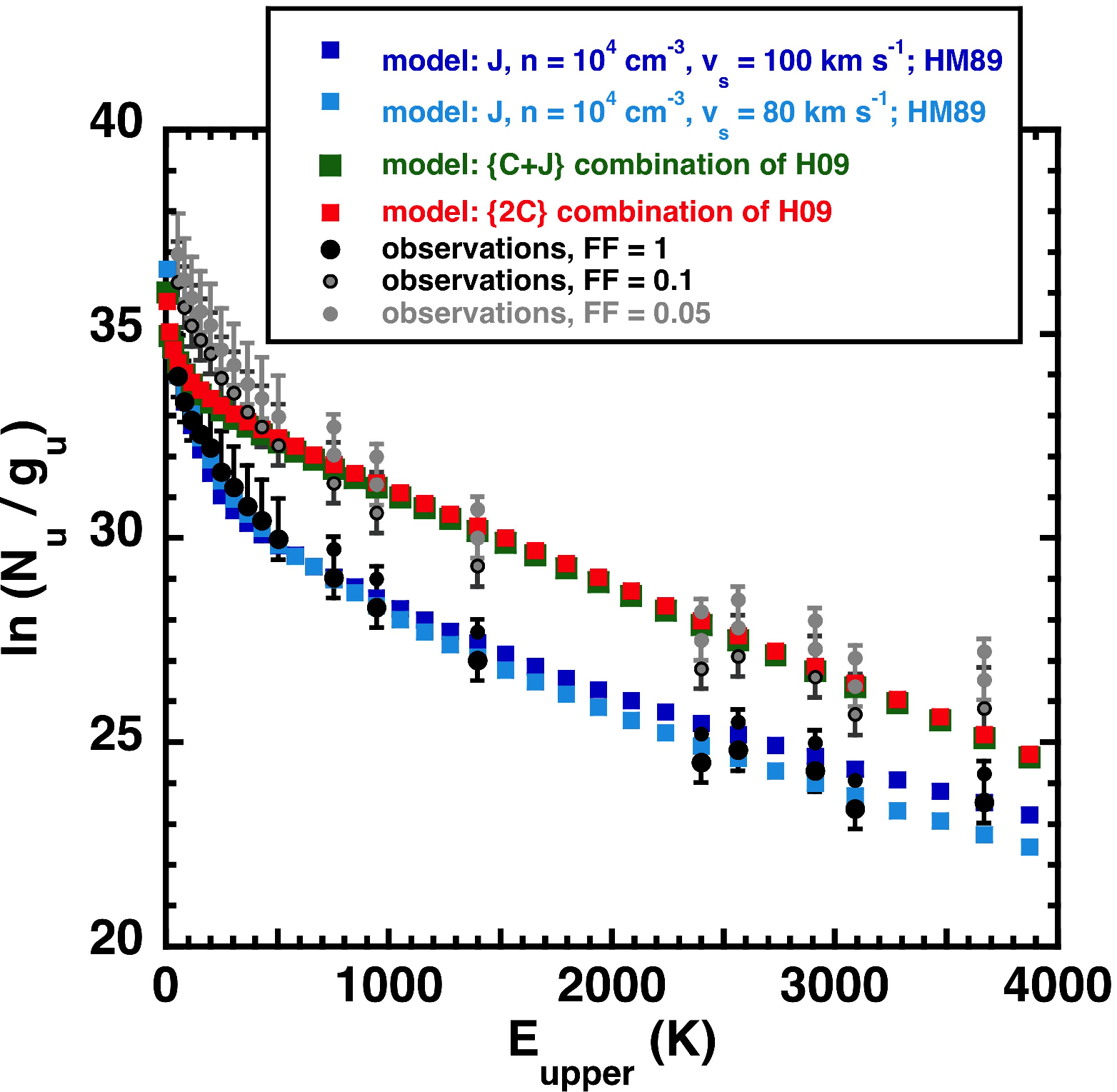}
\caption{High-J CO rotational diagram as a function of rotational level superposed on
two C-shock model from Gusdorf et al. (2012) 
using the model parameters that produce H$_2$ emission \citep[][H09]{hewitt09},
and J-shock model with a density of 10$^4$cm$^{-3}$ and a shock velocity of 80 km s$^{-1}$ 
from Hollenbach \& McKee (1989). $'$FF$'$ indicates the filling factor.} 
\label{coshockmodel}
\end{figure}

We may compute the column density of ortho-H$_2$O following the procedure
outlined by \citet[][see Eq. 1: N$_{H_{2}O}$ $\propto$ $\int Tdv$]{snell05} for 
water in the low-excitation limit
($n<<n_{\rm crit}$). Using an assumed gas density of $5\times 10^5\,\rm
cm^{-3}$ yields water columns of $6.0\times 10^{13}$, $2.0\times 10^{13}$
and $3.6\times 10^{12}\,\rm cm^{-2}$ in the EBWL, BWL and NWL components,
respectively.
An excitation diagram of the four observed water lines
is shown in Figure \ref{waterrotdiag}.
For H$_2$O molecular data{\footnote {see
http://home.strw.leidenuniv.nl/$\sim$moldata/H2O.html}}, we used
\citet{tennyson01} for the energy levels, and \citet{barber06} for
radiative transition rates, and \citet{faure08} for collision rates.
The analysis of the excitation diagram gives similar total H$_2$O columns of a 
few$\times 10^{13}\,\rm cm^{-2}$
obtained from the low temperature component. 
We do not know which of the H$_2$ temperature components
the H$_2$O emission corresponds to, but the resulting water abundances range from $\sim
10^{-8}$ to $\sim 10^{-5}$ with respect to warm and hot H$_2$,
respectively. We note that the column densities derived above are inversely
proportional to the assumed gas density and thus should be interpreted with
some caution. 

\begin{deluxetable}{llll}
\tabletypesize{\scriptsize}
\setlength{\tabcolsep}{0.03in}
\tablewidth{0pt}
\tablecaption{Gaussian velocity components of HIFI water line at 557 GHz}
\label{coproperties}
\tablehead{\colhead{$\Delta$V} & \colhead{V$_o$} & \colhead{$\int Tdv$} & \colhead{Surf. Brightness} \\
\colhead{ (km/s)} & \colhead{ (km/s) } & \colhead{ (K km/s) } & \colhead{(nWm$^{-2}$~sr$^{-1}$)} }
\startdata
 3.7$\pm$0.20&      17.44$\pm$    0.07&   0.53 & 0.17 \\
 27.1$\pm$0.90&      19.00$\pm$     0.27&    2.95 & 0.98\\
 144.0$\pm$3.9&      33.89$\pm$      1.49&    8.82 & 3.30
\enddata
\label{tablehifi}
\end{deluxetable}

\begin{deluxetable}{lrrr}
\tablewidth{0pt}
\tablecaption{Fluxes of PACS-detected H$_2$O and CO lines}
\startdata
\tablehead{   \colhead{Species} & \colhead{Frequency in GHz} & \colhead{E(up)} &
\colhead{S.B.}  \\ 
\colhead{Transition} & \colhead{(Wavelength in $\mu$m)} & \colhead{ (K)} &
\colhead{(nWm$^{-2}$~sr$^{-1}$)} }
o-H$_2$O 2$_{12}$-1$_{01}$ &      1669.90 (179.65)&      114.3&         18.020\\
o-H$_2$O 3$_{03}$-2$_{12}$ &      1716.77 (174.74)&      196.7&         14.335\\
o-H$_2$O 4$_{14}$-3$_{03}$ &      2640.47 (113.61)&      323.4&         39.543\\
CO (4-3)&      461.04 (      650.70)&      55.3&       7.503\\
CO (5-4)&      576.26 (      520.59)&      82.9&       12.35\\
CO (6-5)&      691.47 (      433.85)&      116.1&       19.42\\
CO (7-6)&      806.65  (      371.90)&      154.8&       29.74\\
CO (8-7)&      921.80  (      325.45)&      199.1&       41.56\\
CO (9-8)&      1036.91 (      289.32)&      248.8&       41.31\\
CO (10-9)&     1151.99 (      260.42)&      304.1&      47.87\\
CO (11-10)&    1267.01  (      236.77)&      364.9&     47.87\\
CO (12-11)&    1382.00    (      217.07)&      431.3&     51.86\\
CO (13-12)&    1496.92   (      200.41)&      503.1&     48.79\\

CO (16-15)&      1841.35 (      162.92)&      751.7&     53.13\\
CO (18-17)&      2070.68  (      144.88)&      944.9&     45.53\\
CO (22-21)&      2528.17  (      118.66)&      1397.4&     32.62\\
CO (29-28)&      3325.01   (      90.225)&      2399.8&     9.713\\
CO (30-29)&      3438.36  (      87.250)&      2564.8&     15.29\\
CO (32-31)&      3664.68  (      81.862)&      2911.2&     12.26\\
CO (33-32)&      3777.64 (      79.414)&      3092.5&     5.614\\
CO (36-35)&      4115.61 (      72.893)&      3668.8&     9.595\\
$[O~I]$     &      2061.04 (145.557)   & 188      & 500.9   \\
$[C~II]$    &      1901.58 (157.763)   & 92    & 735.6    
\enddata
\label{tablepacs}
\end{deluxetable}

\begin{figure}[!h]
\plotone{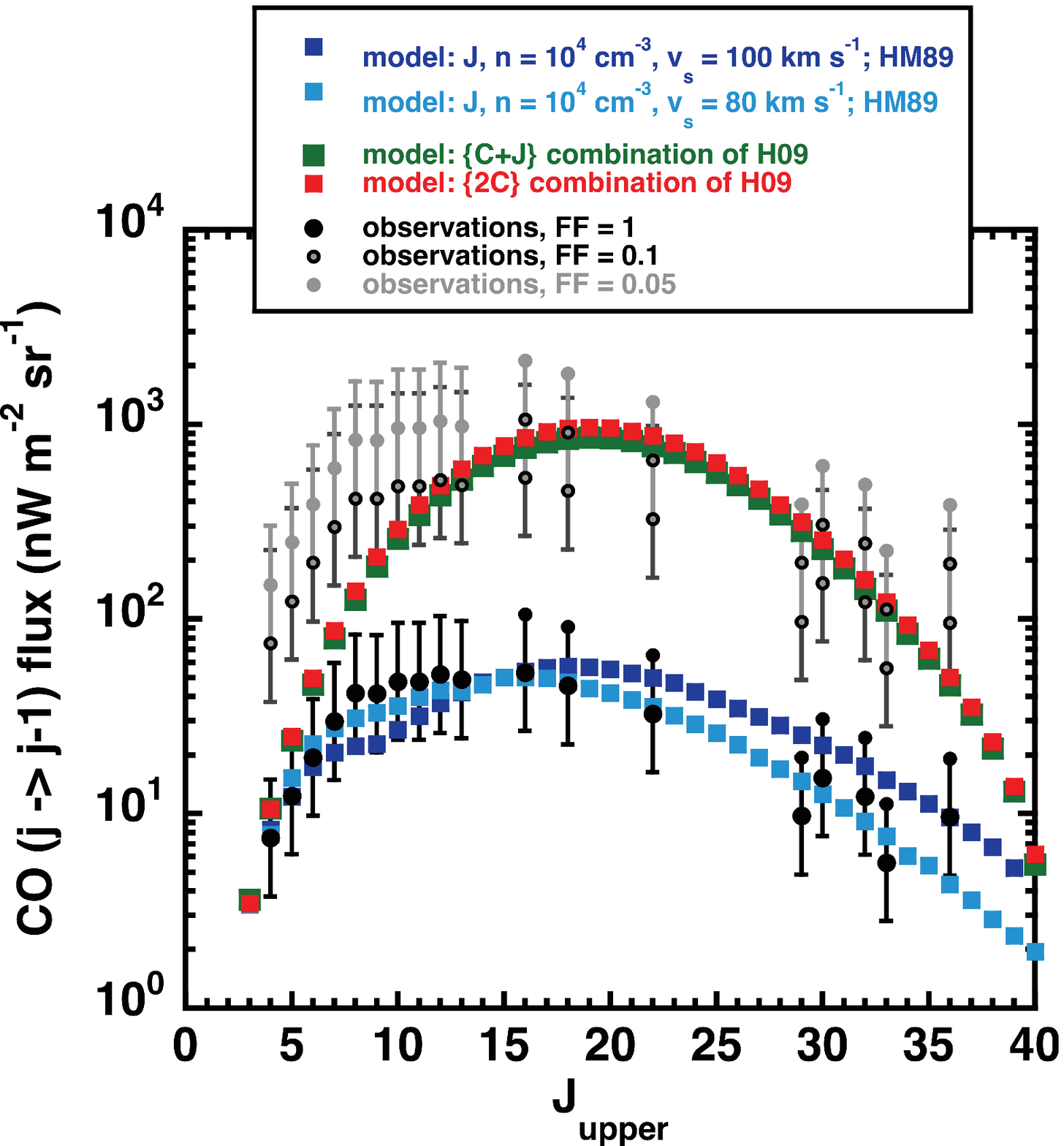}
\caption{High-J CO brightness diagram as a function of rotational level superposed on
two C-shock model from Gusdorf et al. (2012)
using the model parameters that produce H$_2$ emission \citep[][H09]{hewitt09},
and J-shock model with a density of 10$^4$cm$^{-3}$ and a shock velocity of 80 km s$^{-1}$
from Hollenbach \& McKee (1989). $'$FF$'$ indicates a filling factor.}
\label{cosbshockmodel}
\end{figure}

\begin{figure}[!h]
\plotone{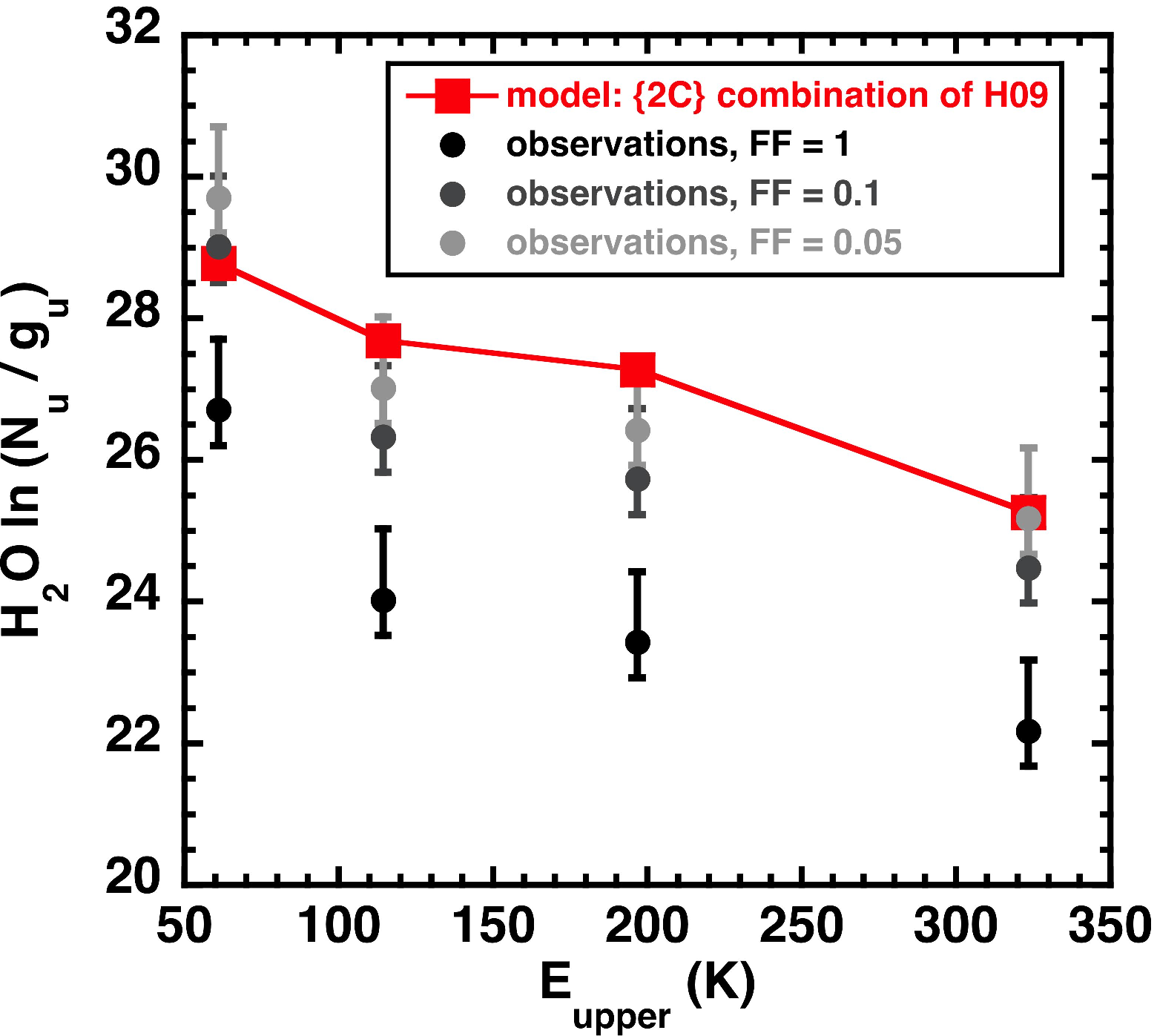}
\caption{H$_2$O rotational diagram as a function of energy level for the same C-shock model as that in Figures 
\ref{coshockmodel} and \ref{cosbshockmodel}.} 
\label{watercshockmodel}
\end{figure}

We compare the H$_2$O, CO and H$_2$ brightnesses with shock models. Bright
H$_2$ emission was detected with the \spitzer\ IRS, and \citet{hewitt09}
showed two possible scenarios of shock models based on H$_2$ emission. 
Their best-fit model is a combination of two C-shock models, one with a pre-shock
density n$_0$ = 10$^6$ cm$^{-3}$ and a shock velocity V$_s$ = 10 \kms, and
the other with n$_0$ = 10$^5$ cm$^{-3}$ with V$_s$ = 40 \kms. The associated filling
factors ($\Phi$ or FF) are 0.61 and 0.007, respectively. The second best
model is a combination of a C-shock (with a density of n$_0$ =
10$^6$ cm$^{-3}$  and V$_s$ = 10 \kms)  and a J-shock 
(with a n$_0$ = 10$^4$ cm$^{-3}$ and 
V$_s$ = 110 \kms). First, we tried the C-shock model following
the method presented in Gusdorf et al. (2008, 2012) to compare the CO data
with shock models. This model solves the magneto-hydrodynamical equations
in parallel with a large chemical network for stationary C- and J-type
shocks (Flower et al. 2003) and combine the outputs with a radiative
transfer module based on the LVG approximation. 
A combination of the two C-shock model 
that reproduced the H$_2$ emission with FF$\sim$1 (see Fig. 7
from Hewitt et al. 2009) can not re-produce either CO (Fig.
\ref{coshockmodel} or \ref{cosbshockmodel}) or H$_2$O (Fig.
\ref{watercshockmodel}). 
The same C-shock model using FF of 0.05 may
reproduce high-J CO lines, but low-J lines don't match with the model.
The same C-shock model could not produce the observed
H$_2$O lines with an assumption of FF=1; however, it does reasonably match the H$_2$O lines 
with an assumption of FF=0.05
(see Fig. \ref{watercshockmodel}).

We find a J-shock  model with a velocity of 80 -100  \kms\ and a
density of n$_0$ $\sim$ 10$^4$ cm$^{-3}$ (HM89) provides a good fit to
the CO rotational diagram (see Fig. \ref{coshockmodel}), and is
consistent with the EBWL flux. A direct comparison of
the surface brightnesses of CO with the J-type shock models of HM89
(their Fig. 7) is shown in Figure \ref{cosbshockmodel}. The
total intensity of the observed water lines are 76.4 nW m$^{-2}$
sr$^{-1}$, which is comparable to the predicted flux from the J-shock
model with velocity of 80 -100  \kms\ and a
density of n$_0$ $\sim$ 10$^4$ cm$^{-3}$ (see Fig. 9 of HM89).
This J-shock model can also explain the
fluxes of [O~I] 145 $\mu$m and [C~II] 157 $\mu$m observed with the
Herschel PACS and [Ne II] and [Ne III] lines with {\it Spitzer} IRS.
It is surprising that the same J-shock model
can explain both CO and ionic lines, because {\it Spitzer} results show that the molecular (H$_2$)
lines segregate from ionic lines within the {\it Spitzer} IRS slits
(H09). Based on {\it Spitzer} data, a principal component
analysis of the spectral-line maps by \citet{neufeld07} also shows
that molecular lines (H$_2$ J$>$2) belong different group from ionic
lines and can be explained by different shock model from that for
ionic lines. If a blast wave with a velocity of 80 km s$^{-1}$ moves
into a dense clump, it would send a converging spherical shock into
the clump. The blue- and red- shifted lines could be shocked clumps
coming towards us or moving away from us with 80 \kms\ velocity,
respectively; this causes  a broad line as large as 160 \kms. In
contrast, water emission in another molecular interacting SNR IC 443
shows the same width as those in CO, 20 - 30 \kms, which show
the ambiguity between J-shock and C-shock models as discussed in
\citet{snell05}. The EBWL in G349.7+0.2 is a direct evidence of
high-velocity J-shocked material. The same J-shock could not produce
the observed H$_2$ brightnesses of G349.7+0.2. \citet{hewitt09}
suggested that H$_2$ emission is explained with the C-shock model.

The observed water gas may or may not be in LTE because these densities are
much lower than the critical density of water ($\sim$10$^9$ cm$^{-3}$). We
have run grids of non-LTE RADEX \citep{vandertak07} models for density varying
from 10$^4$ to 10$^7$ cm$^{-3}$ and column density of H$_2$O varying
from 3$\times$$10^{11}$ to 3$\times$10$^{15}$ cm$^{-2}$ for temperature
from 25 K to 500 K, with steps of 25 K. A water column density of 10$^{13}$
- 10$^{14}$ cm$^{-2}$ and the inferred density of  10$^{4}$ cm$^{-3}$ -
10$^{6}$ cm$^{-3}$ is consistent with the four water lines and the kinetic
temperature is greater than the excitation temperature obtained from the
LTE model (see Fig. \ref{waterrotdiag}). The predicted
fluxes from non-LTE models with a column density of 3$\times$10$^{13}$
cm$^{-2}$ and a density of 10$^{5}$ cm$^{-3}$ for a kinetic temperature
range of 100 - 1000 K are also shown in Figure \ref{waterrotdiag}. The
fluxes from the model are consistent with the observed line brightnesses.
Even if the lines are thick, the transitions are
effectively thin (see detailed discussion in S05) because the photons leak out by sequential
absorption and re-radiation without collisional de-excitation.

We compared the abundances of H$_2$O/H$_2$ and H$_2$O/CO using the
water column density obtained from the non-LTE model described above.
For a temperature of $\sim$200 K, the abundance of H$_2$O/H$_2$ is
approximately 10$^{-6}$ - 10$^{-7}$ using non-LTE value of N(H$_2$O)
of 10$^{13-14}$ cm$^{-2}$; the column density derived from LTE model
could underestimate the abundance. The abundance of H$_2$O/CO is 0.1 -
0.003 for a range of temperature of 150-600 K. The abundances are
comparable to those estimated for 3C391 using ISO \citep{reach98}.
There are two reasons that it is challenging to determine if our
estimated abundance is low or high compared with shock models or other objects. 
We show
a J-shock model with a high velocity ($\sim$ 80 km s$^{-1}$) is
consistent with the rotational diagram of CO emission and the total
flux  of water. (Detailed J-shock models of water lines with high
shock velocities ($>$50 km s$^{-1}$) are currently not available.) 
Most water abundance comparisons have been done with C-shock models
\citep{kaufman96, flower10}. 
Moreover, our water lines show
three components (VBL, BL and NL) while we don't have resolved CO
lines. For our analysis we assume that all PACS lines have the
same profiles as that of the HIFI water line. The resolved CO line is required 
to accurately estimate the H$_2$O/CO abundance
since the
portion of VBL, BL and NL to the total may be different. One 
would need to estimate the abundance for each of VBL, BL and NL,
respectively. 
Behind a J-shock, the abundance of water rises due to the rapid neutral-neutral reactions; 
later, H$_2$O can freeze out. 
Perhaps it is possible that water in EBWL has a
higher abundance that those of NL or BL components. 

The extremely broad water line could be simply from short-lived
molecules formed in high velocity J-shocks which have not had time to freeze out
However, it may be from more complicated geometry such as
high-velocity water bullets or an expanding shell in high-velocity.
Future observations of velocity-resolved CO,
H$_2$O, OH and H$_2$ would be crucial to advance our understanding of
oxygen chemistry of neutral-neutral and/or ion-neutral reactions in
the process of converting oxygen to H$_2$O.

{\vskip 0.2truecm}
Support for this work, part of the NASA Herschel Science Center
(through JPL/Caltech), Astrophysics Data Analysis Program (grant
NNX12AG97G) and Theoretical Research/Laboratory Astrophysics Program was provided by NASA. We thanks David Hollenbach for
insightful and critical comments which helped to significantly improve
the paper. We thank Tom Pannuti for discussion on SNRs,  Herschel
Science Center staff members including Ivan Valtchanov, David
Teyssier, and David Shupe for their support on data reduction, and
calibration, and Emmanuel Caux for various discussion on and updating
CASSIS.

{}

\end{document}